\documentclass[twocolumn]{aastex62}

\def\cii      {[C\thinspace {\scriptsize II}]}
\def\oiiil   {[O\thinspace {\scriptsize III}]$_{88}$}
\def\oiii     {[O\thinspace {\scriptsize III}]}

\usepackage{natbib}

\graphicspath{{./}{figures/}}

\shorttitle{[\ion{O}{3}] emission in a Quasar + Companion at z\,$\sim$\,6}
\shortauthors{Walter et al.}

\begin{document}

\title{\Large \bf No evidence for enhanced \oiii\ 88$\mu$m emission in a z$\sim$6 quasar compared to its companion starbursting galaxy}

\correspondingauthor{Fabian Walter}
\email{walter@mpia.de}

\author[0000-0003-4793-7880]{Fabian Walter}
\affil{Max Planck Institute for Astronomy, K\"onigstuhl 17, 69117 Heidelberg, Germany}
\affil{National Radio Astronomy Observatory, Pete V. Domenici Array Science Center, P.O. Box O, Socorro, NM 87801, USA}

\author[0000-0001-9585-1462]{Dominik Riechers}
\affiliation{Cornell University, 220 Space Sciences Building, Ithaca, NY 14853, USA}
\affil{Max Planck Institute for Astronomy, K\"onigstuhl 17, 69117 Heidelberg, Germany}

\author{Mladen Novak}
\affiliation{Max Planck Institute for Astronomy, K\"onigstuhl 17, 69117 Heidelberg, Germany}

\author[0000-0002-2662-8803]{Roberto Decarli}
\affiliation{INAF—Osservatorio di Astrofisica e Scienza dello Spazio, via Gobetti 93/3, I-40129, Bologna, Italy}

\author{Carl Ferkinhoff}
\affiliation{Winona State University, Winona, MN 55987, USA}

\author[0000-0001-9024-8322]{Bram Venemans}
\affil{Max Planck Institute for Astronomy, K\"onigstuhl 17, 69117 Heidelberg, Germany}

\author[0000-0002-2931-7824]{Eduardo Ba\~nados}
\affiliation{The Observatories of the Carnegie Institution for Science, 813 Santa Barbara Street, Pasadena, CA 91101, USA}

\author{Frank Bertoldi}
\affil{Argelander-Institut f\"ur Astronomie, Universit\"at Bonn, Auf dem H\"ugel 71, 53121 Bonn, Germany}

\author{Chris Carilli}
\affil{National Radio Astronomy Observatory, Pete V. Domenici Array Science Center, P.O. Box O, Socorro, NM 87801, USA}

\author{Xiaohui Fan}
\affil{Steward Observatory, The University of Arizona, 933 N.\ Cherry Ave., Tucson, AZ 85721, USA}

\author{Emanuele Farina}
\affil{University of California Santa Barbara, Santa Barbara, CA, USA}

\author{Chiara Mazzucchelli}
\affil{European Southern Observatory, Alonso de C\'ordova 3107, Vitacura, Regi\'on Metropolitana, Chile}

\author[0000-0002-9838-8191]{Marcel Neeleman}
\affil{Max Planck Institute for Astronomy, K\"onigstuhl 17, 69117 Heidelberg, Germany}

\author{Hans--Walter Rix}
\affil{Max Planck Institute for Astronomy, K\"onigstuhl 17, 69117 Heidelberg, Germany}

\author{Michael Strauss}
\affil{Department of Astrophysical Sciences, Princeton University, Princeton, NJ 08544 USA}

\author{Bade Uzgil}
\affil{National Radio Astronomy Observatory, Pete V. Domenici Array Science Center, P.O. Box O, Socorro, NM 87801, USA}
\affil{Max Planck Institute for Astronomy, K\"onigstuhl 17, 69117 Heidelberg, Germany}

\author{Ran Wang}
\affil{Kavli Institute of Astronomy and Astrophysics at Peking University, No.5 Yiheyuan Road, Haidian District, Beijing, 100871, China}




\begin{abstract}

We present ALMA band~8 observations of the \oiii\ 88$\mu$m line and the underlying thermal infrared continuum emission in the z=6.08 quasar CFHQS~J2100--1715 and its dust--obscured starburst companion galaxy (projected distance: $\sim$60\,kpc). Each galaxy hosts dust--obscured star formation at rates $>$\,100\,M$_\odot$\,yr$^{-1}$, but only the quasar shows evidence for an accreting 10$^9$\,M$_\odot$ black hole. Therefore we can compare the properties of the interstellar medium in distinct galactic environments in two physically associated objects, $\sim$1\,Gyr after the Big Bang. Bright \oiii\ 88$\mu$m emission from ionized gas is detected in both systems;  the positions and line--widths are consistent with earlier \cii\ measurements, indicating that both lines trace the same gravitational potential  on galactic scales. The \oiii\ 88$\mu$m/FIR luminosity ratios in both sources fall in the upper range observed in local luminous infrared galaxies of similar dust temperature, although the ratio of the quasar is smaller than in the companion. This suggests that gas ionization by the quasar (expected to lead to strong optical \oiii\ 5008\,\AA\ emission) does not dominantly determine the quasar's FIR \oiii\ 88$\mu$m luminosity. Both the inferred number of  photons needed for the creation of O$^{++}$ and the typical line ratios can be accounted for without invoking extreme (top--heavy) stellar initial mass functions in the starbursts of both sources.

\end{abstract}

\keywords{galaxies: high-redshift; galaxies: ISM; quasars: emission lines; quasars: general}


\section{Introduction} \label{sec:intro}

The most distant quasars at z$>$6 are unique probes of galaxy and structure formation in the first Gigayear of the universe. Their rapidly accreting supermassive black holes make these quasars the most luminous (non--transient) sources known at this cosmic epoch \citep[e.g.,][]{fan06,venemans15,banados16,wu15}.
The interstellar medium of  more than 30 host galaxies of z$\sim$6 quasars has now been detected  through the \cii\ 158 $\mu$m line (hereafter: \cii) and far--infrared (FIR) continuum emission. This suggests that the host galaxies have intense star formation rates at many 100's of M$_\odot$yr$^{-1}$ occuring coevally with the growth of the accreting central black holes \citep[e.g.,][]{bertoldi03,walter03,walter09,wang13,willott13,willott15,decarli18,venemans18}.

Using ALMA observations, \cite{decarli17} found dust--obscured companions to 4 quasars in a \cii\ survey of 27\,z$\sim$\,6 quasars. These massive star-forming companion galaxies show no signs of AGN activity and were detected serendipitously within 8–-60 kpc at the same redshifts in four systems, suggesting that they are physically associated with overdensities or proto--clusters of galaxies at z$>$6. These companion galaxies rival the FIR and \cii\ luminosities of the neighboring quasars, in some cases even surpassing their (already extreme) luminosity. These quasar--galaxy pairs thus provide a unique opportunity to efficiently study the interstellar medium in distinctly different galactic environments in the first Gyr of the Universe: pure starbursts, and those in the presence of a powerful AGN.

\begin{figure}
\includegraphics[scale=.37,angle=0]{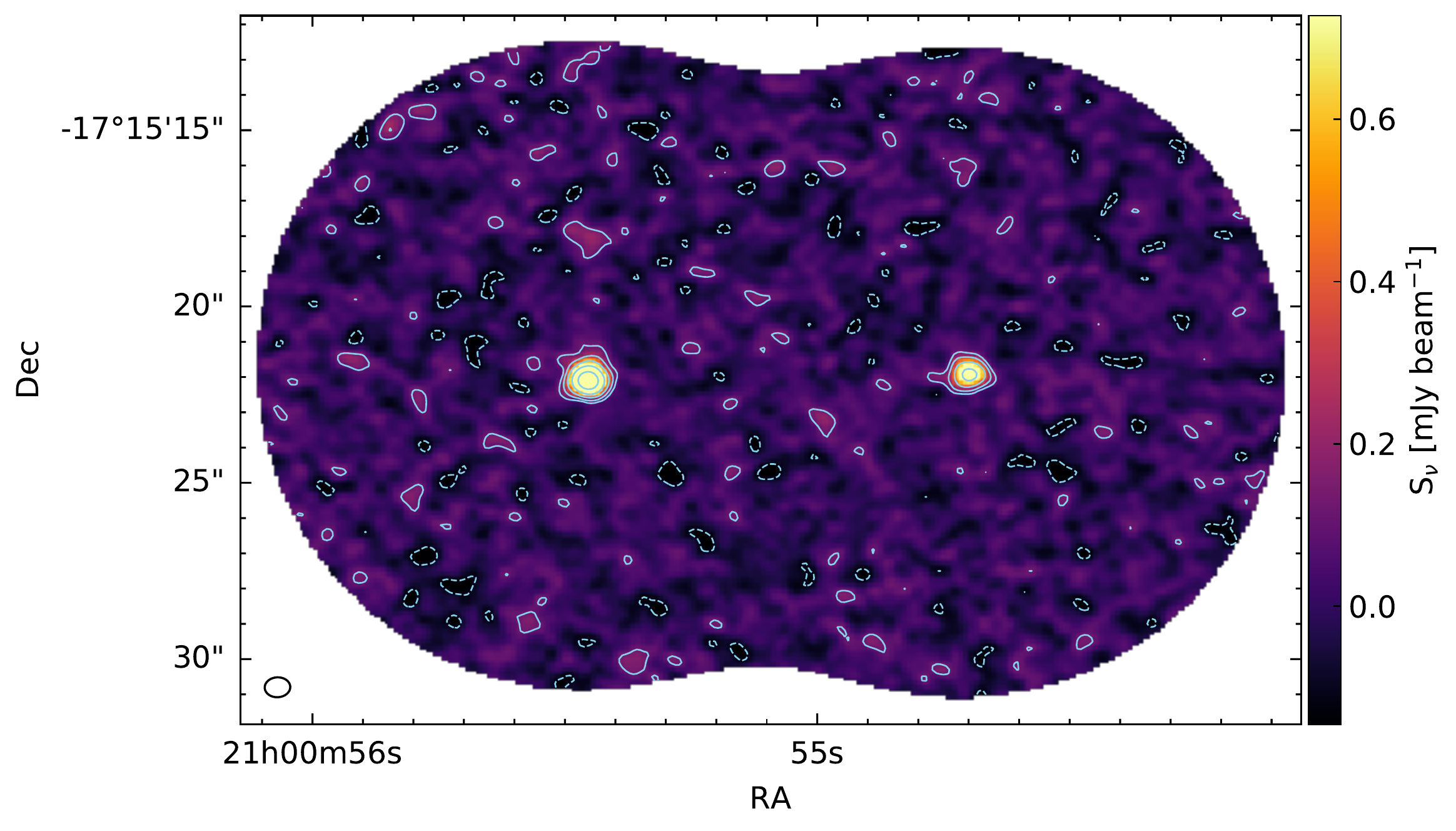}
\caption{Restframe 88$\mu$m continuum emission of the z=6.08 quasar J2100--Q (Western source) and its companion J2100--SB (Eastern source). Logarithmically spaced contours are shown at $\pm$2, 4, 8, ... $\sigma$, with $\sigma$=54\,$\mu$Jy\,beam$^{-1}$. The beamsize of 0.73$"\times$0.57$"$ (position angle: 87.3$^\circ$) is shown in the bottom left corner. Two pointings were needed to cover both sources in ALMA band~8.\label{fig:continuum}}
\end{figure}

To date, the \cii\ line has been the workhorse line of ISM studies  at z$>$6 \citep[e.g.,][]{carilliwalter13}, but it is only one of several fine structure lines that help to charaterize the physical properties of the interstellar medium \citep[e.g.,][]{diaz-santos17}. Of these, the \oiii\ 88\,$\mu$m line ($^3$P$_1\rightarrow^3$P$_0$) (hereafter: \oiiil), emitted by doubly ionized oxygen (O$^{++}$), is a particularly important diagnostic of the star formation process \citep[e.g.,][]{ferkinhoff10,ferkinhoff11,ferkinhoff15,vishwas18}: Given the O$^+$ ionization potential of 35.1\,eV, O$^{++}$ is almost exclusively found in dense \ion{H}{2} regions around O--type stars or in AGN environments. Indeed, an enhancement of the optical 5008\AA\ \oiii\ line is a key diagnostic of AGN activity \citep[e.g.,][]{baldwin81}. In contrast, neutral carbon has an ionization potential of only 11.3\,eV, and \cii\ is therefore be emitted both in the neutral and ionized phase. So far, \oiiil\ detections have been reported in a number of z\,$>$\,6 galaxies \citep[e.g.,][]{inoue16,laporte17,carniani17,tamura18,hashimoto18a,hashimoto18b,marrone18}. However, in this redshift range, only 2 systems have been detected in \oiiil, \cii, {\em and} the underlying dust continuum: the Lyman break galaxy B14--65666 at z=7.15 \citep[]{hashimoto18b} and the lensed galaxy SPT0311--58 at z=6.90 \citep[the latter actually consists of two sources]{marrone18}.

In this paper we report \oiiil\ and underlying dust continuum observations with ALMA of the z=6.08 quasar CFHQS\,J2100--1715 (hereafter J2100--Q) and its dust--enshrouded companion starburst (J2100--SB). J2100--Q was discovered by \cite{willott10a}, and a black hole mass of 9.4$\pm$2.6$\times$10$^8$\,M$_\odot$ was reported in \cite{willott10b}. J2100--SB was discovered by \cite{decarli17} in the far--infrared using ALMA. Multi--wavelength follow--up observations of the companion galaxy using Spitzer (3.6$\mu$m, 4.5$\mu$m) and HST (WFC3/140W) as well as MUSE and LBT spectroscopy did not detect the companion in the rest--frame UV/optical, with an implied ratio of obscured to un--obscured star formation of $>$99\% (Mazzucchelli et al. {\em in prep.}).  No metallicity measurement of our targets exists, but they must be significantly enriched with heavy elements as they harbor significant amounts of dust, and have high ($\sim$100\,M$_\odot$\,yr$^{-1}$) star formation rates. 

In Section~2 we describe our ALMA band~8 observations, in Sec.~3 we present our results, followed by a discussion and a summary in Secs.~4 and~5. Throughout this paper we use cosmological parameters H$_0$\,=\,70 km\,s$^{-1}$ Mpc$^{−1}$, $\Omega_{\rm M}$\,=\,0.3, and $\Omega_{\rm \Lambda}$\,=\,0.7, in agreement with Planck Collaboration XVI (2014), leading to a scale of 5.67 kpc per arcsec at the redshift of our sources. 

\section{Observations}

We observed the \oiiil\ line ($\nu_{\rm rest}$\,=\,3393.0062\,GHz) of J2100--Q and J2100--SB redshifted to $\nu_{\rm obs}$=479.2GHz at z=6.08 with ALMA in band~8 and configuration~C43--1 on 2018~May~20 (two executions) and 2018--May--23 (one execution) for a total of 3$\times$5200\,s or 4.3\,h, of which 2.5\,h were spent on source. The redshifts of the sources were known from the earlier detections of [\cii] line emission (z=6.0806$\pm$0.0011 for the quasar, 6.0796$\pm$0.0008 for the companion). The two sources are very close in redshift ($\Delta v_{\rm LOS}$=41\,km\,s$^{-1}$, \citealt{decarli17}), meaning that the \oiiil\ emission in the two objects could be covered with the same frequency setup. However, the projected separation on the sky ($\sim$11$\arcsec$) required two pointings to cover the sources in band~8. 
The software package CASA \citep{mcmullin07} was used for data reduction and imaging. For optimal sensitivity, we employed natural weighting when imaging the data, leading to a synthesized beam size of 0.73$"\times$0.56$"$ ($\sim$3.7\,kpc at z=6.08), and an rms of 0.45\,mJy\,beam$^{-1}$ per 50\,km/s (80\,MHz) channel. The corresponding sensitivity in the continuum (excluding the channels that contain line emission) is 54\,$\mu$Jy\,beam$^{-1}$ over an effective bandwidth of 5.625\,GHz. For all analysis of the \oiiil\ line data, the continuum has been subtracted in the {\em uv}--plane using line--free channels. 

We also re--analyze the \cii\ and underlying continuum ALMA data published by \cite{decarli17} and \cite{venemans18}. The beam size of these \cii\ data (0.73$''\times$0.57$''$) is almost identical to that presented here. Therefore, no additional beam matching was required. In order to optimize the S/N in the measurements, and given the fact that any extent, if present, is marginal, we  extract \cii\ line and underlying continuum fluxes at the peak position of both sources. We report the \cii\ values adopted in this study in Tab.~1.\footnote{We note that both \cite{decarli17} and \cite{venemans18} report \cii\ and FIR luminosities that are slightly higher than the values adopted here, as their reported fluxes are based on 2D fitting of the line emission in the image domain.}

\begin{figure}
\begin{center}
\includegraphics[scale=.45,angle=0]{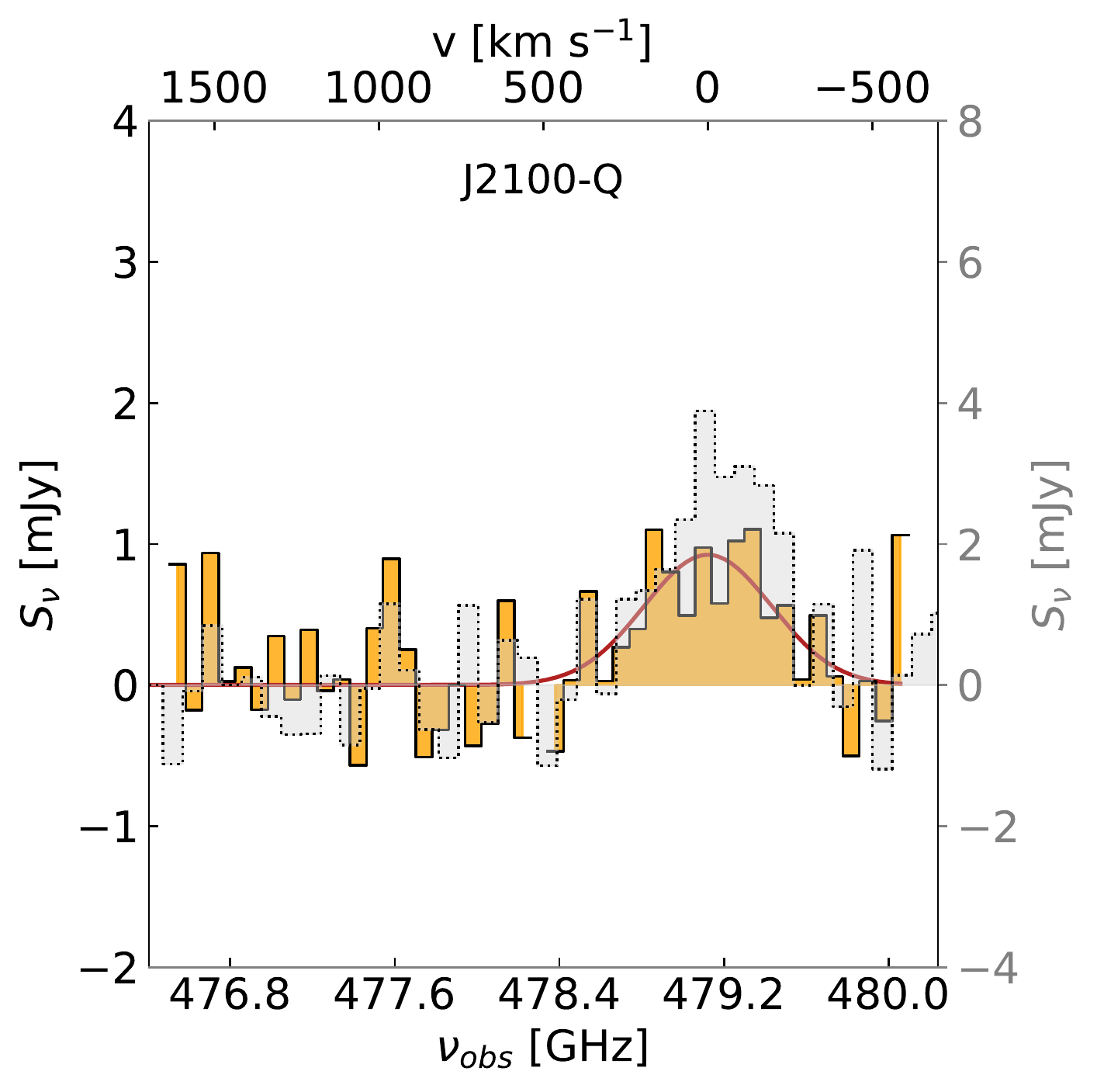}
\includegraphics[scale=.45,angle=0]{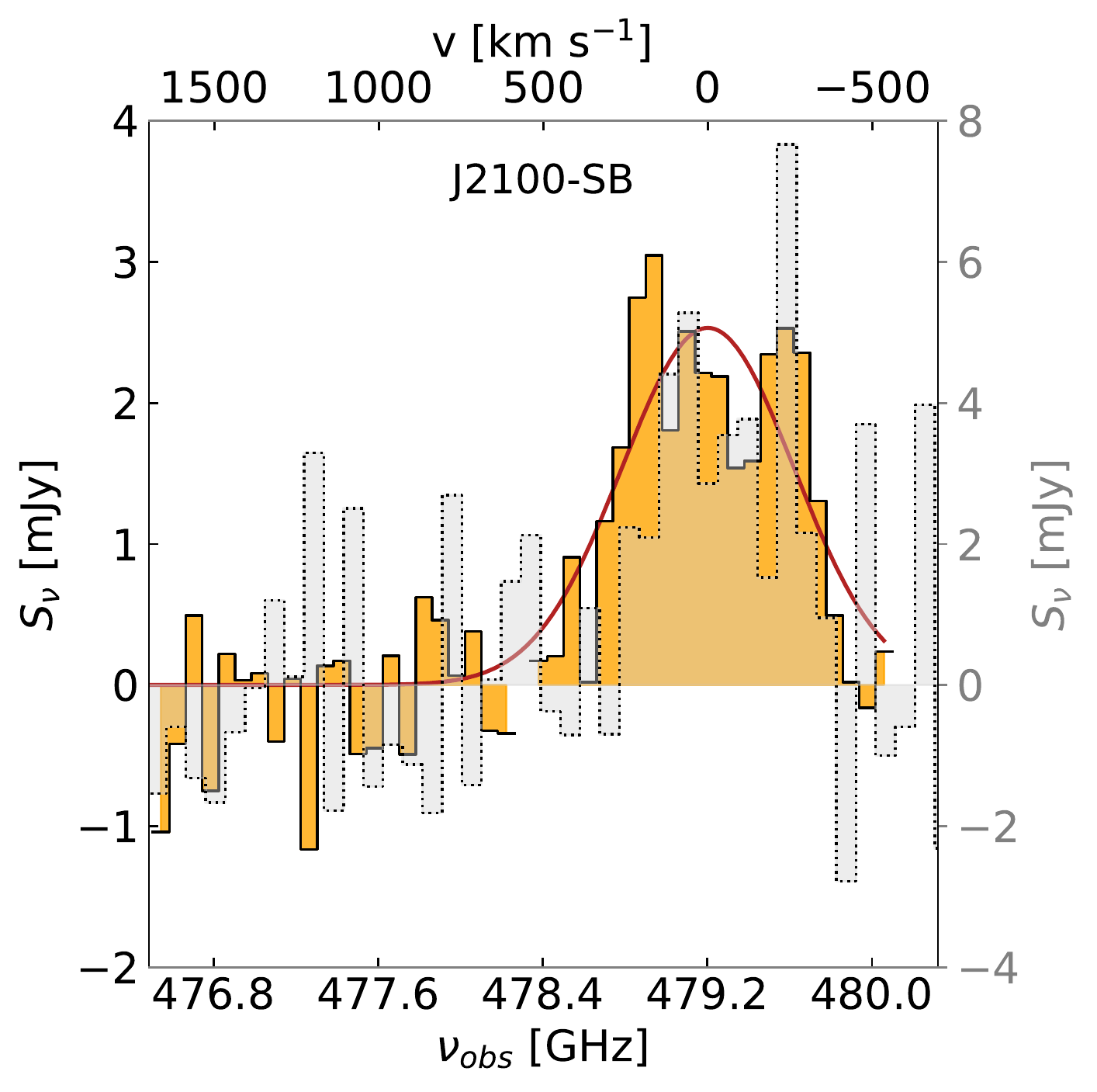}
\end{center}
\caption{\oiiil\ emission line spectra of J2100--Q (top) and J2100--SB (bottom) after continuum subtraction. The red line shows the Gaussian fit (parameters reported in Tab.~1). In greyscale, we overplot the respective spectra of the \cii\ line (after shifting to the same rest--frame velocity as the \oiiil), again after continuum subtraction. The \cii\ spectrum has been scaled down in intensity by a factor of 2 (axis labels to the right). 
\label{fig:spectra}}
\end{figure}

\section{Results}

\subsection{Continuum emission}

In Figure~1 we show the rest--frame 88\,$\mu$m (observed: 475\,GHz) continuum map of J2100--Q and J2100--SB. This map is based on those channels in the data cube that do not contain the \oiiil\ line. The sources are detected at high significance and are unresolved at the resolution of our beam (0.73$"\times$0.57$"$, flux densities are reported in Tab.~1). The coordinates (RA/DEC in J2000.0) of the sources, 21:00:54.70, --17:15:21.9 (J2100--Q), and 21:00:55.45, --17:15:21.7 (J2100--SB) are in agreement with \cite{decarli17}. 

Together with our earlier restframe 158\,$\mu$m continuum measurement around the \cii\ line (\citealt{decarli17}, see Tab.~1 for slightly updated numbers) we can use the new restframe 88\,$\mu$m observations to further constrain the properties of the FIR continuum emission. The IR and FIR luminosities  are obtained by fitting the 88\,$\mu$m and 158\,$\mu$m data points (assuming they are optically thin) with a modified blackbody, taking the effect of the cosmic microwave background into account \citep{dacunha13}. These results are reported in Tab.~1. The resulting L$_{\rm IR}$--based star formation rates \citep{kennicutt12} are 125$\pm$14 M$_\odot$\,yr$^{-1}$ (J2100--Q) and 284$\pm$30 M$_\odot$\,yr$^{-1}$ (J2100--SB).

\subsection{\oiiil\ line emission}

We show the continuum--subtracted \oiiil\ spectra of both sources in Fig.~2 as extracted from the peak positions of the continuum map. The Gaussian fit results are reported in Tab.~1. For reference, we overplot scaled, continuum--subtracted \cii\ spectra from \cite{decarli17} in greyscale. Both the redshifts and the linewidths of the \oiiil\ and \cii\ detections agree within the uncertainties.  The \oiiil\ line width of the quasar is much smaller than that of emission lines seen in the broad line region of the quasar (FWHM: $\sim$3600\,km\,s$^{-1}$, \citealt{willott10b}).
In Fig.~3 we show the integrated \oiiil\ maps of both sources, after continuum subtraction. 
2D--Gaussian fitting shows that the \oiiil\ line emission is unresolved at our resolution like the continuum emission. In Tab.~1 we summarize the resulting \oiiil\ luminosities. Using the \oiiil--SFR scaling relations of \cite{delooze14} for their `high--redshift sample' we derive \oiiil--based star formation rates of approximately 100\,M$_\odot$\,yr$^{-1}$ and 360\,M$_\odot$\,yr$^{-1}$ for the quasar and the companion, respectively, consistent with the FIR--based estimates (Sec.~3.1).

\begin{deluxetable}{cccc}
\tablecaption{\oiiil\ and rest--frame 88\,$\mu$m continuum measurements of the quasar CFHQS J2100--Q and its companion J2100--SB. Re--measured \cii\ and restframe 158\,$\mu$m continuum values from the \cite{decarli17} data are also given.\label{tab:numbers}}
\tablewidth{0pt}
\tablehead{
\colhead{} & \colhead{units} & \colhead{quasar} & \colhead{companion} }
\startdata
z$_{\rm [OIII]88}$                            &                  & 6.0816$\pm$0.0009 & 6.0806$\pm$0.0005 \\
\hline
S$_{\rm cont, 88\mu m}$\tablenotemark{a}    & mJy\,beam$^{-1}$ & 1.12\,$\pm$\,0.05 & 2.87\,$\pm$\,0.05 \\
S$_{\rm cont, 158\mu m}$\tablenotemark{a,b} & mJy\,beam$^{-1}$ & 0.46\,$\pm$\,0.07 & 1.37\,$\pm$\,0.14 \\
\hline
S$_{\rm [OIII]88}$\tablenotemark{c}    & mJy\,beam$^{-1}$ & 0.93\,$\pm$\,0.20  & 2.55\,$\pm$\,0.21  \\
FWHM$_{\rm [OIII]88}$\tablenotemark{c} & km\,s$^{-1}$     & 454\,$\pm$\,117    & 614\,$\pm$\,62  \\
F$_{\rm [OIII]88}$\tablenotemark{d}    & Jy\,km\,s$^{-1}$ & 0.39\,$\pm$\,0.06  & 1.52\,$\pm$\,0.07 \\
F$_{\rm [CII]}$\tablenotemark{d}     & Jy\,km\,s$^{-1}$ & 1.09\,$\pm$\,0.08  & 1.89\,$\pm$\,0.21  \\
\hline
T$_{\rm dust}$                                 & K                   & 41$\pm$1.2                & 37$\pm$1.2  \\
L$_{\rm IR[8-1000\mu m]}$                      & 10$^{11}$ L$_\odot$ & 8.3\,$\pm$\,0.37  & 19.1\,$\pm$\,0.33 \\
L$_{\rm FIR[42-122\mu m]}$                     & 10$^{11}$ L$_\odot$ & 6.8\,$\pm$\,0.30 & 15.5\,$\pm$\,0.28 \\
\hline
L$_{\rm [OIII]88}$                  & 10$^9$ L$_\odot$    & 0.77 $\pm$ 0.12        & 2.86 $\pm$ 0.13\\
L$_{\rm [CII]}$                 & 10$^9$ L$_\odot$    & 1.05 $\pm$ 0.08            & 1.81 $\pm$ 0.9\\
\hline
L$_{\rm [OIII]88}$/L$_{\rm [CII]}$  &                    & 0.74 $\pm$ 0.15            & 1.58 $\pm$ 0.24\\
L$_{\rm [OIII]88}$/L$_{\rm FIR}$    &    10$^{-3}$                & 1.12 $\pm$ 0.11   & 1.86 $\pm$ 0.14 \\
L$_{\rm [CII]}$/L$_{\rm FIR}$     &      10$^{-3}$                & 1.55 $\pm$ 0.13   & 1.17 $\pm$ 0.13  \\
\enddata
\tablecomments{All measurements obtained at continuum emission peak.}
\tablenotetext{a}{Observed continuum flux density underlying the line emission.} 
\tablenotetext{b}{The error of the companion is higher as it is not in the phase centre.}
\tablenotetext{c}{From Gaussian fitting to the line.}
\tablenotetext{d}{Based on integrated line maps (values consistent with spectra.}
\end{deluxetable}

\begin{figure}
\begin{center}
\includegraphics[scale=.4,angle=0]{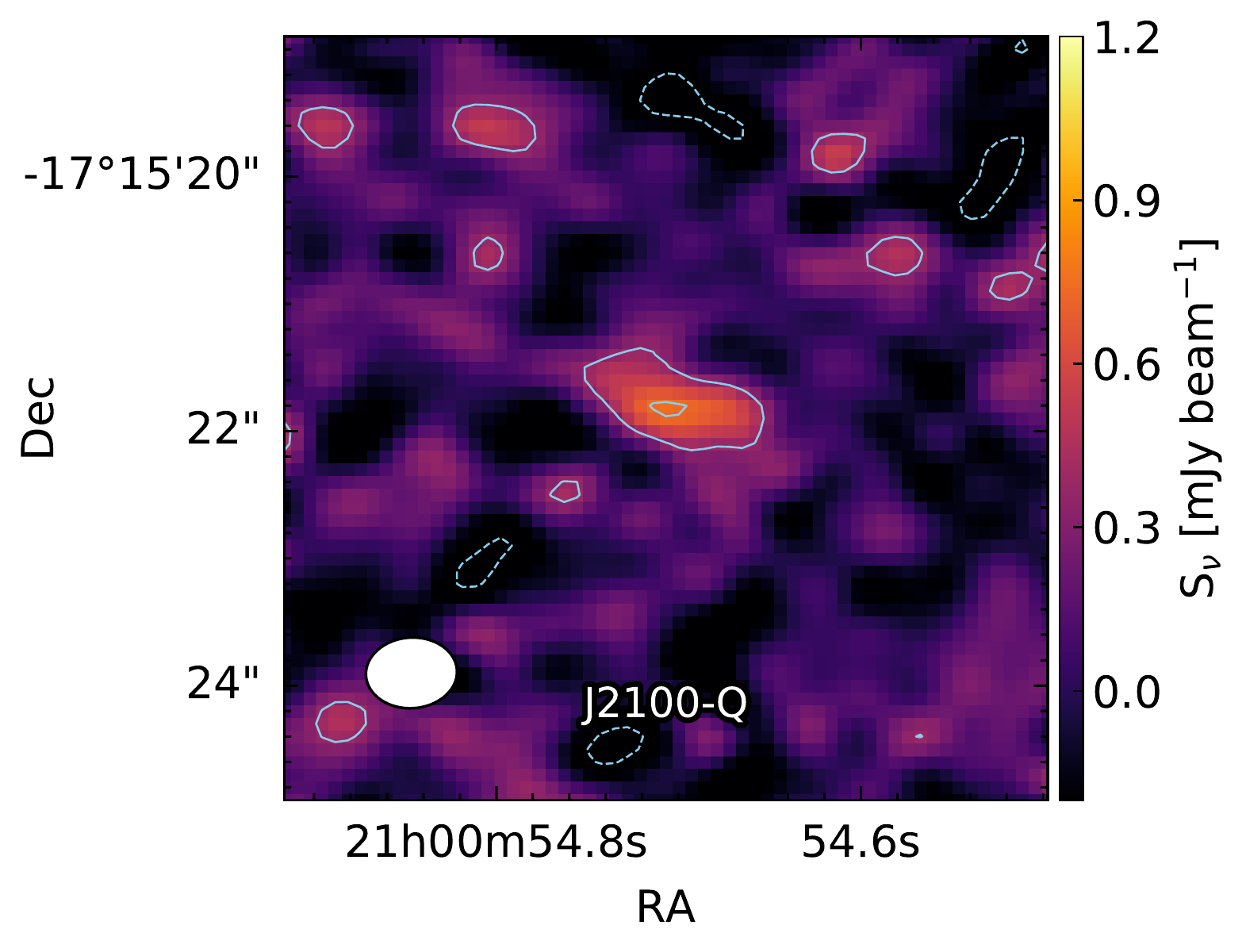}\\
\includegraphics[scale=.4,angle=0]{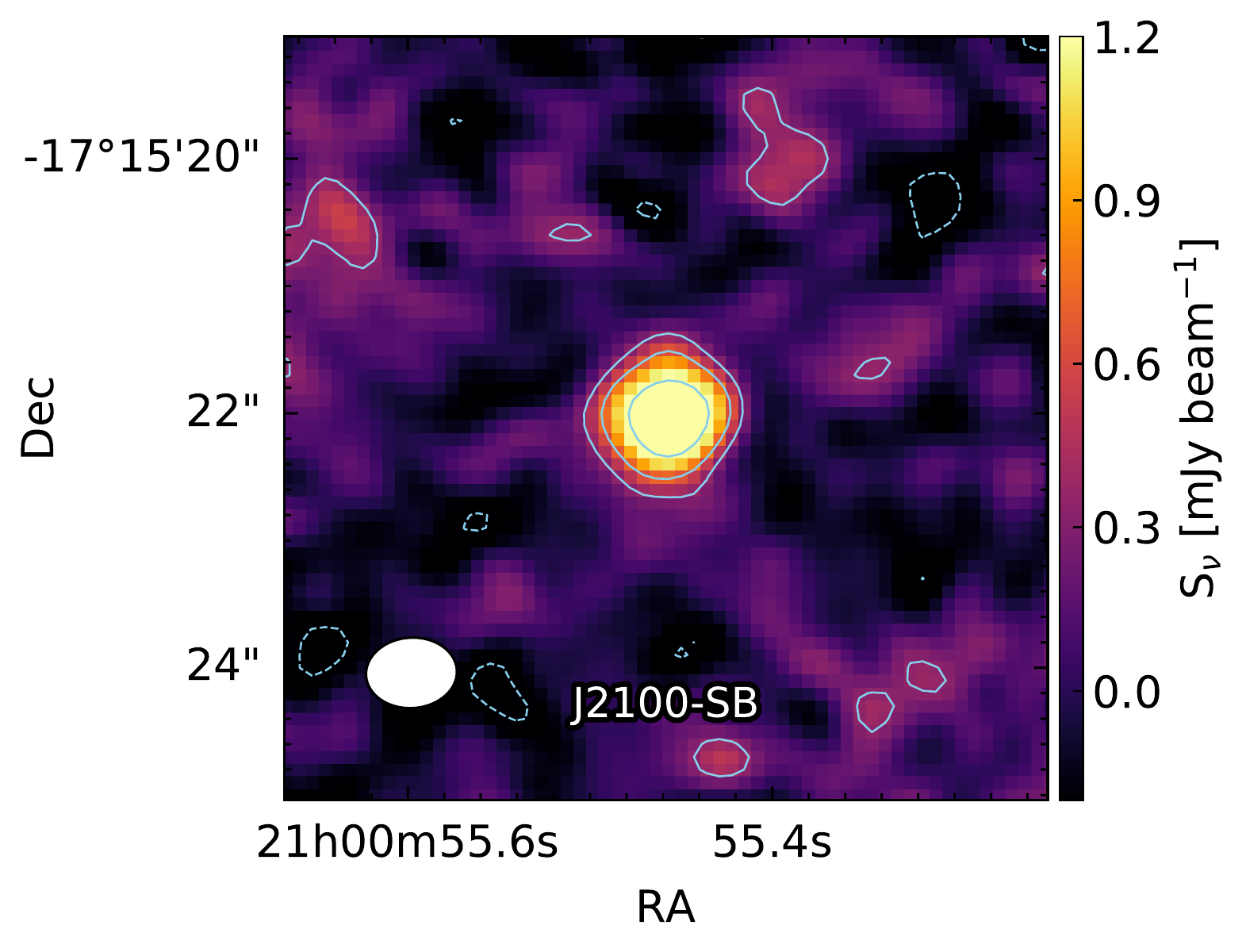}
\end{center}
\caption{\oiiil\ emission after continuum subtraction in J2100--Q (top) and J2100--SB (bottom), integrated over a velocity range of 1.2 $\times$\,FWHM, respectively. Logarithmically spaced contours are shown at $\pm$2, 4, 8 $\sigma$, with $\sigma$=0.17\,mJy\,beam$^{-1}$ for J2100--Q and $\sigma$=0.16\,mJy\,beam$^{-1}$ for J2100--SB. The beamsize of 0.73$"\times$0.57$"$ (position angle: 87.3$^\circ$) is shown in the bottom left corner.
\label{fig:oiii}}
\end{figure}

\begin{figure*}
\includegraphics[width=6cm]{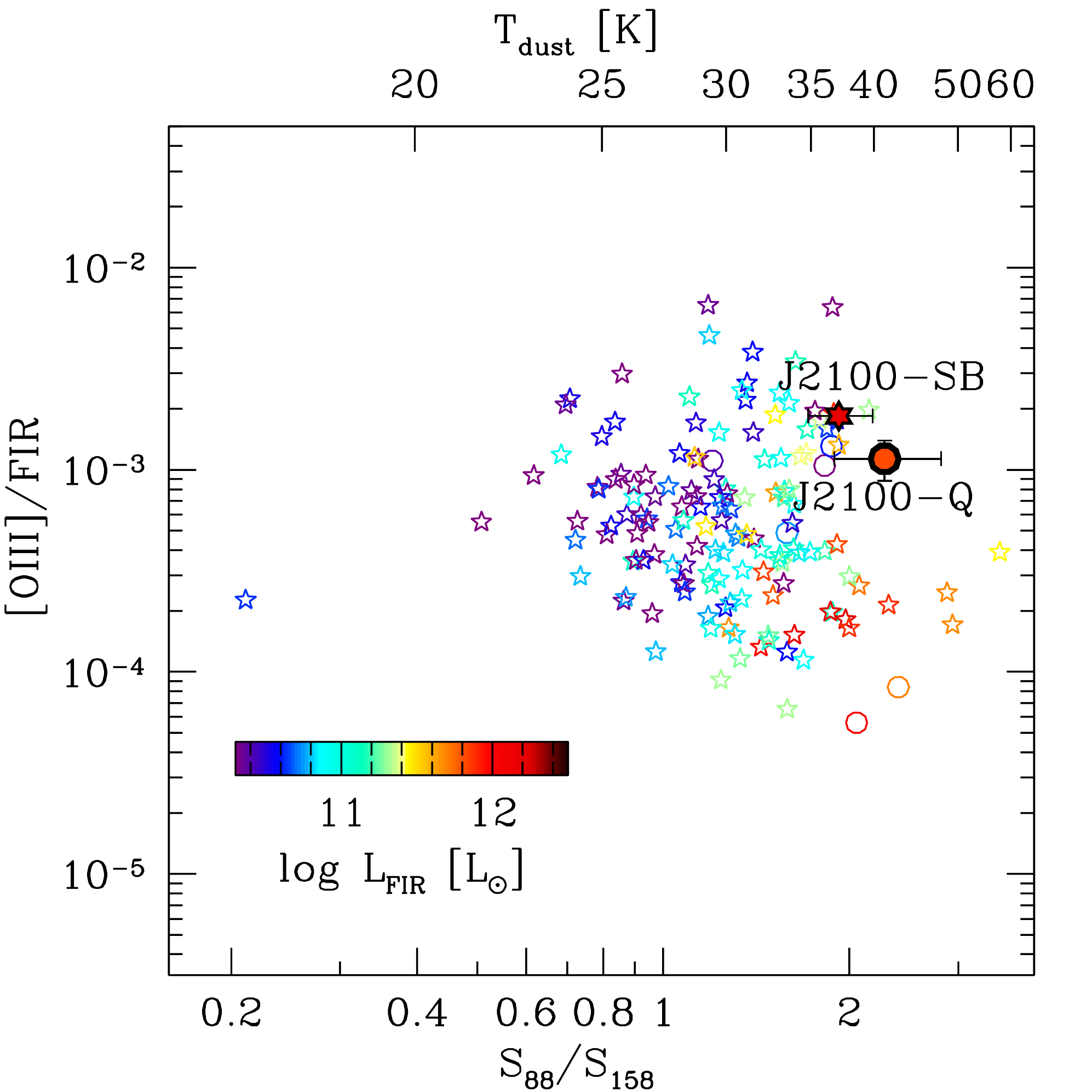}\includegraphics[width=6cm]{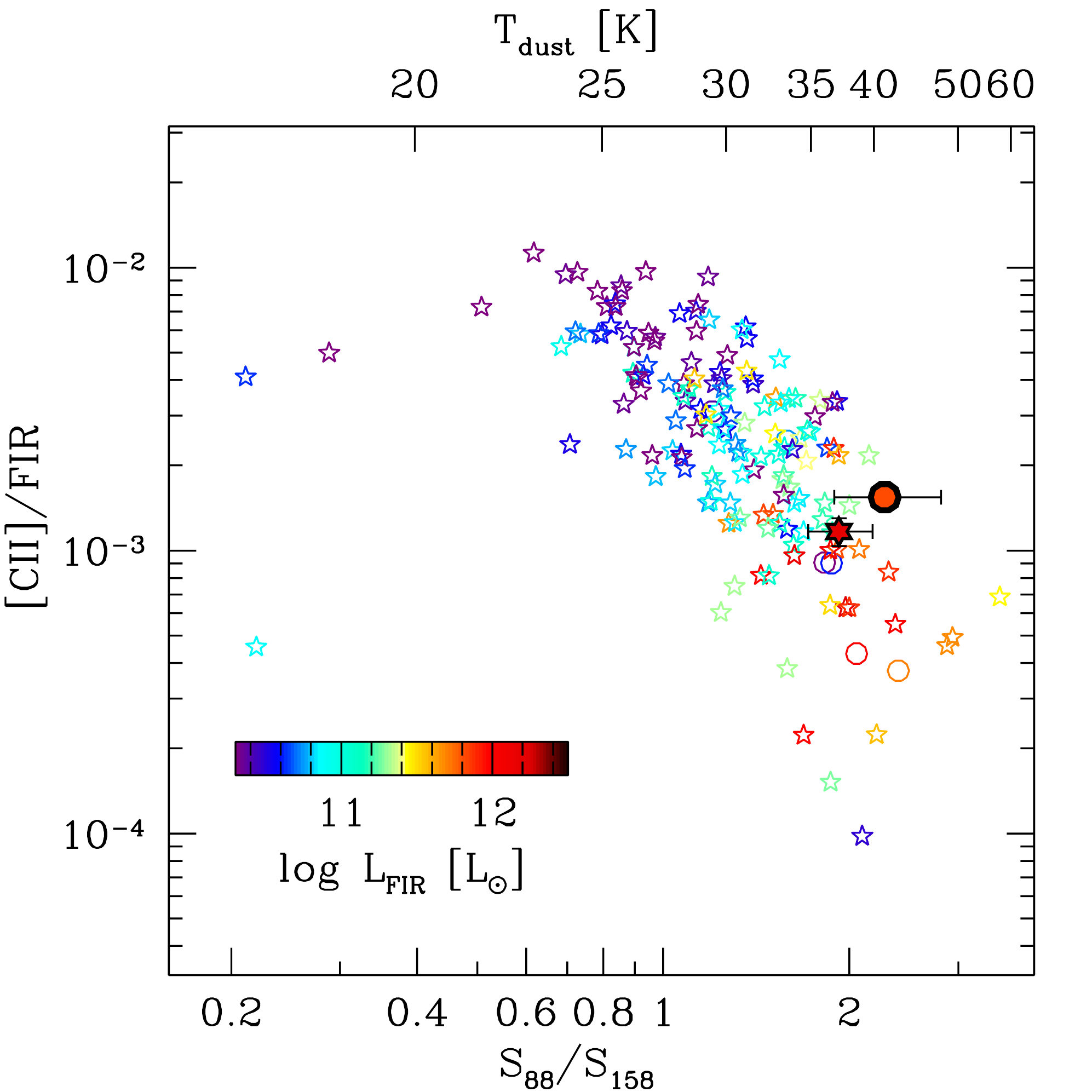}\includegraphics[width=6cm]{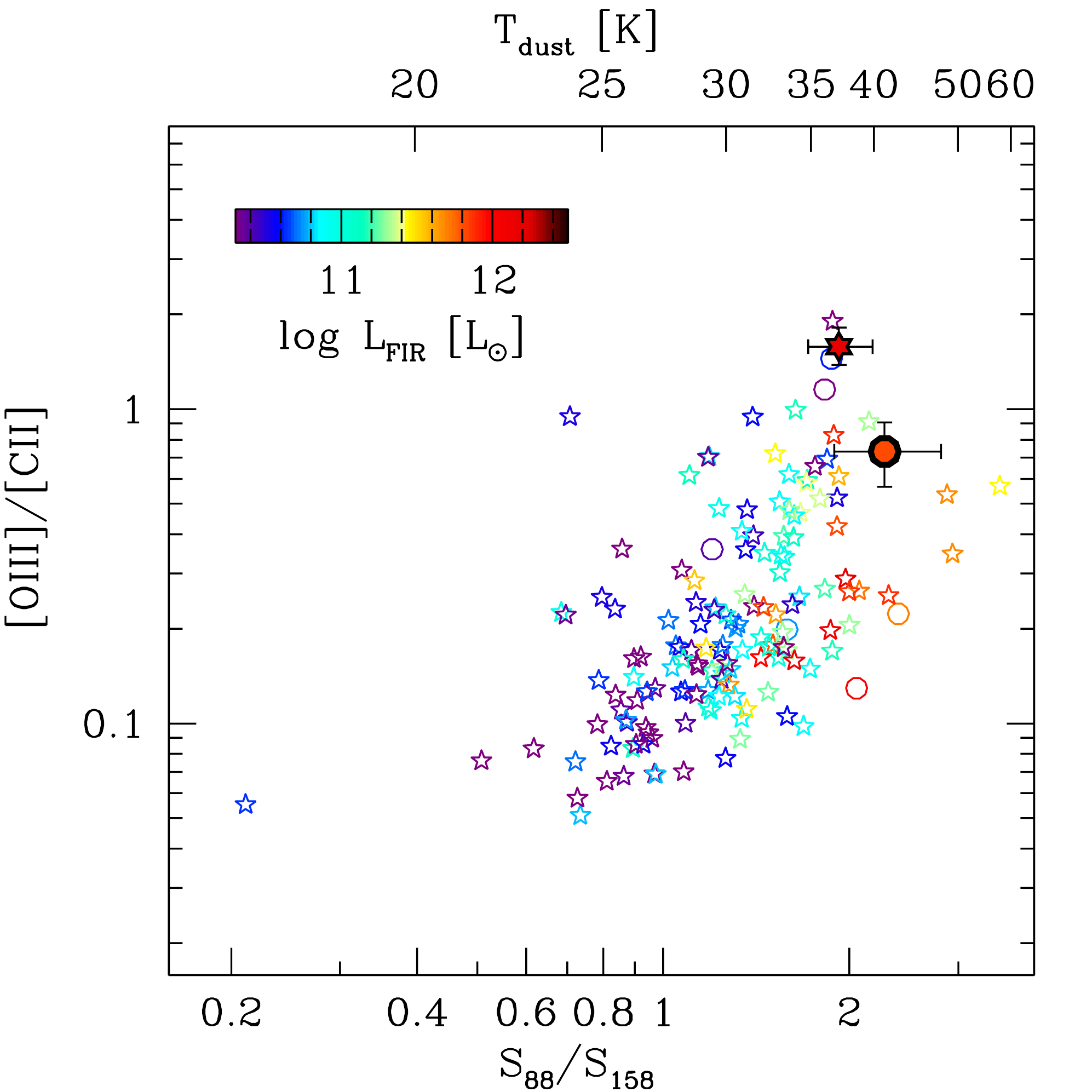}
\caption{Line/FIR ratios of the z=6.08 quasar J2100--Q and its companion J2100--SB compared to the local LIRG compilation by \cite{diaz-santos17}. The data points are color--coded by FIR luminosity, circles (stars) mark sources with an AGN contribution of $>$50\% ($<$50\%) in the MIR \citep{diaz-santos17}. {\em Left panel:} \oiiil/FIR luminosity ratio as a function of the continuum flux density ratio (S$_{88 \mu m}$/S$_{158 \mu m}$). This ratio is a proxy for the temperature as shown on the top x--axis, assuming  a modified black body with $\beta=1.6$ for the dust emissivity. {\em Middle and right panels:} Same x--axis as in the left panel but now showing the \cii/FIR luminosity ratio and the \oiii/\cii\  ratio, respectively.
}
\end{figure*}

\subsection{Line/FIR luminosity ratio}

We now combine the \oiiil\ and rest--frame 88$\mu$m continuum measurements with the earlier \cii\ line measurements to constrain the properties of the interstellar medium in our targets. This is particularly interesting, as our sources are physically quite different. In Fig.~4 (left panel) we plot the \oiiil/FIR luminosity ratios vs.\  the continuum flux density ratios (S$_{88 \mu m}$/S$_{158 \mu m}$) of our sources, the latter being a proxy for the temperature of the dust (color--coded by FIR luminosity), and compare them to the recent compilation of local luminous infrared galaxies (LIRGs) by \cite{diaz-santos17}. The mid--infrared emission in the majority of the local sample is not dominated by AGN emission (circles vs.\ stars in Fig.~4). 

Our two sources have luminosities at the high end of the LIRG sample studied in \cite{diaz-santos17}, but their \oiiil/FIR luminosity ratios lie only slightly above the average LIRG value of similar dust temperature, within the scatter found locally. Perhaps unexpectedly, despite the presence of an accreting 10$^{9}$\,M$_\odot$, the quasar does not show enhanced \oiiil emission compared to either the companion or the local comparison sample.

It is instructive to also compare the \cii/FIR luminosity ratios for our sources to the local LIRGs (middle panel of Fig.~4). The \cii/FIR luminosity ratios of our targets ($\sim$1.6$\times$10$^{-3}$) are located  within the distribution observed for the local LIRGS at the same temperature (around $\sim$1$\times$10$^{-3}$). The right panel in Fig.~4 shows the \oiiil/\cii\ luminosity ratios as a function of dust temperature. Again, our two targets fall within the distribution seen in local LIRGS, and the \oiii/\cii\ ratio is actually {\em lower} in the quasar compared to the companion.

\section{Discussion}

\subsection{Origin of the \cii\ and \oiiil\ emission}

The line--widths and positions of the \oiii\ and \cii\ emission are identical within the uncertainties, which indicates that they trace the same gravitational potential  on galactic scales (cf.\ \citealt{carniani17}).
\cite{diaz-santos17} show that the ratio of the \cii\ emission emerging from PDRs, to that from ionized gas, is correlated with S$_{63 \mu m}$/S$_{158 \mu m}$ (see their Fig.~3). We determine the S$_{63 \mu m}$/S$_{158 \mu m}$ ratio from our S$_{88 \mu m}$/S$_{158 \mu m}$ measurements assuming the same blackbody parameters as used in Sec.~3.1 (see also Tab.~1), finding S$_{63 \mu m}$/S$_{158 \mu m}$\,=\,1.7 for the quasar, and 1.8 for the companion, respectively.

If the correlations of \cite{diaz-santos17} hold, then more than 80\% of the \cii\ emission in both the quasar and the companion are from the neutral phase of the interstellar medium (L$_{\rm [CII]}^{\rm ionized}$\,$<$\,0.2\,L$_{\rm [CII]}$). This is in contrast to the \oiiil\ emission, which only traces the ionized phase, because the O$^+$ ionization potential is 35.1\,eV (the next ionization level is at 54.9\,eV, e.g., \citealt{kramida18}). Following \cite{ferkinhoff10,ferkinhoff11} and \cite{vacca96}, 
a blackbody with a temperature $>$35,000\,K ($>$45,000\,K) produces significant emission of photons with energies $> 35$\,eV ($>$55\,eV). These temperatures correspond to stars of type O7V and O3V, respectively. 

The presence of a FUV--bright central source (i.e., the accreting black hole in J2100--Q) will will lead to optical 5008\,\AA\ \oiii\ line emission in the broad line region (e.g., \citealt{vandenberg01}) and the narrow line region (e.g., \citealt{baldwin81}) that will become accessible to the {\em James Webb Space Telescope} at z\,=\,6.08.  The FIR \oiiil\ line on the other hand is likely to arise from stellar \ion{H}{2}\ regions given that it has the same systemic velocity and line width as the \cii\ line, and is much narrower than the optical lines emerging from the BLR. Additionally, the \oiiil/FIR and \oiiil/\cii\ ratios are similar to those seen for systems where the emission is known to arise from \ion{H}{2} regions. 

\subsection{Minimum ionized gas mass}

We can infer the mass of doubly--ionized oxygen in our targets from the observed \oiiil\ luminosity \citep[see, e.g.,][]{ferkinhoff11} to be $M$(O$^{++}$)/M$_\odot$=$2.55\times10^{-4}$ $L_{\rm [OIII]88}$/L$_\odot$ = $1.7\times10^{5}$\,M$_\odot$ for the quasar, and $7.3\times10^{5}$\,M$_\odot$ for the companion. Assuming that most the oxygen is in doubly--ionized form, and that the oxygen abundance is $\chi$(O/H)=$5.9\times10^{-4}$ \citep{savage96}, this implies a minimum mass of ionized gas of $M_{\rm min}$(H$^+$)=$2.1\times10^7$\,M$_\odot$ for the J2100--Q, and $M_{\rm min}$(H$^+$)=$7.7\times10^7$\,M$_\odot$ for J2100--SB.  Other high-z systems have shown O$^{++}$/O fractions of $\sim$10\% \citep{ferkinhoff11}, so that the actual ionized gas mass could be a factor of 10 higher than our minimum value here. Using the dust mass estimates of 3.2$\times$10$^8$\,M$_\odot$ and 5.5$\times$10$^8$\,M$_\odot$ derived for J2100--Q and J2100--SB, respectively \citep{decarli17}, and assuming a standard gas--to--dust ratio of $\sim$100 \citep{draine07, sandstrom13}, the molecular gas content (M(H$_2$)) in both sources is $>$10$^{10}$\,M$_\odot$. A ratio of $M_{min}(H^+)$/M(H$_2$) $\sim$ 10$^{-3}$ is similar to what is found in nearby galaxies (\citealt{brauher08}, see also the compilation in \citealt{ferkinhoff11}, their figure 3). The total ionized gas mass is thus only a small fraction of the total mass of the dense interstellar medium (and thus baryonic mass) in the two objects. 

\subsection{Minimum number of ionizing photons and stars}

Following \cite{ferkinhoff10,ferkinhoff11,ferkinhoff15} and \cite{vishwas18}, we calculate  minimum O$^{+}$--ionizing photon rates for J2100--Q and the J2100--SB of  $Q_0=(1.24 \pm 0.27)\times 10^{54}$\,s$^{-1}$ and $Q_0=(4.62 \pm 0.20)\times 10^{54}$\,s$^{-1}$, respectively. It is interesting to compare this number with back-of-the-envelope expectations based on the observed SFR. The companion galaxy has a SFR$\approx$300\,M$_\odot$\,yr$^{-1}$. The stars that provide significant radiation energetic enough to doubly--ionize oxygen live only for $\lesssim$5\,Myr. Assuming that the SFR is constant, we infer the total stellar mass created in the last 5\,Myr, and distribute it over different stellar mass bins based on a \cite{chabrier03} Initial Mass Function (IMF). The stellar main sequence provides us with an estimate of the luminosity and effective temperature of stars (assuming a simple blackbody) based on their mass, thus we can compute for each stellar mass bin how many photons are produced with energy $h\nu$$>$35.1\,eV, needed to ionize \ion{O}{2}. By integrating over the stellar mass distribution, scaled to match the total mass of stars produced in 5\,Myr, we obtain $Q_{\rm 0, exp}=2.6\times10^{55}$\,s$^{-1}$ (this number changes to $Q_{\rm 0, exp}=2.3\times10^{55}$\,s$^{-1}$ for a \citealt{kroupa01} IMF). This number is slightly higher than the one inferred from our \oiii\ observations (we get similar results for J2100--Q). This suggests that star formation with a standard IMF is sufficient to account for the observed \oiii\ luminosity, with no need to invoke a top-heavy IMF or a contribution by a `buried' quasar.

\section{Summary}

We present ALMA \oiiil\ observations of the z=6.08 quasar J2100--Q and its dust--enshrouded starburst companion J2100--SB. This system is unique, as it offers the possibility to study the physical properties of the interstellar medium of two starbursts that are physically associated with each other within 1\,Gyr of the Big Bang. They display distinct galactic environments: one source shows the presence of a powerful AGN, while the other does not. Interestingly, we find that the \oiiil/FIR luminosity ratio of the starburst companion J2100--SB is higher than that of the quasar J2100--Q, even though the accreting supermassive black hole in the latter provides additional photons with energies high enough to produce O$^{++}$ (which typically leads to strong optical 5008\,\AA \oiii\ emission from the broad and narrow line regions, measureable with JWST in the future). The fact that we do not see enhanced FIR \oiiil\ emission in J2100--Q indicates that its \oiiil\ emission is dominated by star formation, and not AGN activity.  This is supported by the finding that the linewidths and positions of the \oiiil\ and \cii\ line are the same in both sources and consistent with previous findings that the interstellar medium properties in distant quasar host galaxies are predominantly powered by star formation \citep[e.g.,][]{leipski14,barnett15,venemans17}. One caveat is that J2100--SB could in principle also host an {\em obscured} accreting supermassive black hole. In that case, the \oiiil\ emission in the companion may not be solely due to star formation. However, the fact that the \oiiil/FIR and \oiiil/\cii\ luminosity ratios of J2100--SB lie  within the range of local LIRGs (including those that are not dominated by an AGN) argues against such a scenario. This latter finding, together with our analysis of the number of photons needed for the creation of O$^{++}$, implies that no extreme (top--heavy) initial stellar mass functions are needed to explain the \oiiil\ luminosity in our sources. 

\acknowledgments
 
We thank the referee for helpful comments that improved our manuscript and Gordon Stacey for useful suggestions. FW, BV, MN, and MaN acknowledge support from the ERC Advanced Grant 740246 (Cosmic\_Gas). DR acknowledges support from the National Science Foundation under grant number AST--1614213. RW acknowledge supports from the National Science Foundation of China (11721303). This paper makes use of the following ALMA data: ADS/JAO.ALMA\#2017.1.00118.S,\\ ADS/JAO.ALMA\#2015.1.01115.S. ALMA is a partnership of ESO (representing its member states), NSF (USA) and NINS (Japan), together with NRC (Canada), NSC and ASIAA (Taiwan), and KASI (Republic of Korea), in cooperation with the Republic of Chile. The Joint ALMA Observatory is operated by ESO, AUI/NRAO and NAOJ. The National Radio Astronomy Observatory is a facility of the National Science Foundation operated under cooperative agreement by Associated Universities, Inc.

\facilities{ALMA}
\software{CASA \citep{mcmullin07}}


\end{document}